\documentclass[12pt]{iopart}

\usepackage{iopams}
\usepackage{graphicx,subfigure}
\usepackage{url}

\renewcommand{\vec}[1]{\boldsymbol{#1}} 

\begin{document}

\bibliographystyle{apalike}

\title[Complementary Observables and Non-Boolean Logic Outside Quantum Physics]{Complementary Observables and Non-Boolean Logic \\ Outside Quantum Physics}

\author{Harald Atmanspacher$^{1,2}$ and Peter beim Graben$^{3,4}$}

\address{$^1$ Collegium Helveticum, University and ETH Zurich, Schmelzbergstr.~25, \\ CH--8092 Zurich, $^2$ Institute for Frontier Areas of Psychology and Mental Health, \\ Wilhelmstr.~3a, D-79098 Freiburg, $^3$ Department of German Studies and Linguistics, Humboldt-Universit\"at zu Berlin, Unter den Linden 6, D--10099 Berlin, $^4$  Bernstein Center for Computational Neuroscience, Philippstr.~13, D--10115   Berlin}
\ead{atmanspacher@collegium.ethz.ch}

\begin{abstract}

The concept of complementarity in combination with a non-Boolean calculus of propositions refers to a pivotal feature of quantum systems which has long been regarded as a key to their distinction from classical systems. But a non-Boolean logic of complementary features may also apply to classical systems, if their states and observables are defined by partitions of a classical state space. If these partitions do not satisfy certain stability criteria, complementary observables and non-Boolean propositional lattices may be the consequence. This is especially the case for non-generating partitions of nonlinear dynamical systems. We show how this can be understood in more detail and indicate some challenging  consequences for systems outside quantum physics, including mental processes.

\end{abstract}

\maketitle

\section{Introduction}

On 16 September 1927, Bohr introduced the concept of complementarity into physics at the International Conference on Physics at Como. In his lecture {\it The Quantum Postulate and the Recent Development of Atomic Theory} Bohr argued that the postulate of an essential discontinuity in atomic processes entails deep consequences for their proper description. Discussing the infamous wave-particle (light-matter) problem in quantum systems, he stated a ``reciprocal relation between the maximum sharpness of definition of the space-time and energy-momentum vectors associated with an individual'' event. More generally, Bohr referred to this reciprocal relation as a ``complementarity of the space-time description and the claims of causality'' (Bohr 1928, p.~582).

In classical physics, waves and particles are considered as mutually incompatible entities. Therefore, if a system is correctly described in terms of one of these entities, the description in terms of the other one must be incorrect. In quantum physics, however, waves and particles are no ontological entities but manifestations of a system under particular (e.g.~empirical) contexts. Although they mutually exclude one another, they are together necessary to describe the system completely. This is the basic deviation from classical thinking that Bohr suggested to cover by the term complementarity.

Today we know that he imported the notion of complementarity from psychology and philosophy (for details see Holton 1970). With this background it is not astonishing that Bohr was always keen to expand the significance of complementarity beyond physics (cf.~Favrholdt 1999). In his {\it Atomic Theory and the Description of Nature} (Bohr 1934, p.~5) he wrote:
\begin{quote}        \small
We are concerned with the recognition of physical laws which lie outside the domain of our ordinary experience and which present difficulties to our accustomed forms of perception.
\end{quote}

And in his article {\it On the notions of causality and complementarity} (Bohr 1948, p.~318) we read:
\begin{quote} \small
Recognition of complementary relationship is not least required in psychology, where the conditions for analysis and synthesis of experience exhibit striking analogy with the situation in atomic physics. In fact, the use of words like ``thoughts'' and ``sentiments'', equally indispensible to illustrate the diversity of psychical experience, pertain to mutually exclusive situations characterized by a different drawing of the line of separation between subject and object. In particular, the place left for the feeling of volition is afforded by the very circumstance that situations where we experience freedom of will are incompatible with psychological situations where causal analysis is reasonably attempted.
\end{quote}

The concept of complementarity has played a central role in various versions of the Copenhagen interpretation of quantum theory which Bohr originally designed together with Heisenberg. But Bohr's attempts to generalize complementarity beyond physics were joined by only few others, notably Pauli and later Wheeler. It should take a long time, until the 1990s, before their speculations started to be turned into concrete projects (see Wang {\it et al.} 2013 for a brief overview).

Although the wave-particle duality was the historical root of complementarity in physics, a modern perspective allows us to understand it as a consequence of different representations of a system. The present view, first formulated by Bernays (1948), states that complementarity can be related to two basic formal features of quantum theory.
 \begin{enumerate}
\item It can be based on {\it non-commutative algebras of observables}, pioneered by Murray and von Neumann in the mid  1930s. Non-commuting observables are the formal core of their incompatibility, and a ``maximal'' form of incompatibility defines complementarity (Raggio and Rieckers 1983).
\item It can be based on {\it non-distributive lattices of propositions}, pioneered by Birkhoff and von Neumann (1936). Non-distributive lattices are the formal core of the non-Boolean nature of complementary propositions.
 \end{enumerate}
It is important to note that both (i) and (ii) are not restricted to physics in principle, and even less so to quantum physics. While propositions about a system are a very general possibility to characterize it, not every proposition can be turned into a formally well-defined observable. For this reason, (ii) offers a wider scope to apply complementarity outside physics than (i).

In the following we discuss how complementary observables and non-Boolean propositions can arise in descriptions of systems and their dynamics that are based on state spaces. A crucial strategy in this approach is the definition of {\it epistemic} states and associated propositions based on partitions of the relevant space. The way in which the partition is constructed decides whether the resulting descriptions are compatible, resp.~Boolean, or not.

\section{Epistemic Descriptions of Classical Dynamical Systems}
\label{class}
Measurements (or observations) require the preparation of a state of the system to be measured (or observed), choices of initial and boundary conditions for this state, and the selection of particular measurement setups. They refer to operationally defined \emph{observables} which can be deliberately chosen by the experimenter (Pauli 1950, Primas 2007).
\subsection{Observables and Partitions}
\label{stateobs}
A classical dynamical system is characterized by the fact that all observables are compatible with each other. However, in general this holds only for a so-called \emph{ontic description} (Atmanspacher 2000) where the state of a system is considered as if it could be characterized precisely as it is. On such an account, the \emph{ontic state} of the system is given by a point $\vec{x}$ in state space $X$. Classical observables are real-valued functions $f: X \to \mathbb{R}$, such that $a = f(\vec{x})$ is the value of observable $f$ in state $\vec{x}$. A family of observables $\vec{f} = \{f_i: X \to \mathbb{R} | 1 \le i \le n \}$ spans one of many possible \emph{observation spaces} $Y = \vec{f}(X)$ (Birkhoff and von Neumann 1936). Only if all functions $f_i$ spanning the observation space $Y$ are injective, their pre-images contain exactly one point $\vec{x} = f_i^{-1}(a_i)$ for all $i$, and can be called \emph{ontic observables}.
By contrast, \emph{epistemic descriptions} acquire significance if at least one observable is not injective. They refer to the knowledge that can be obtained about an ontic state (Atmanspacher 2000) from representing a measurement result as a point in observation space.
\begin{figure}[!t]
\center
\includegraphics[height=4.5cm]{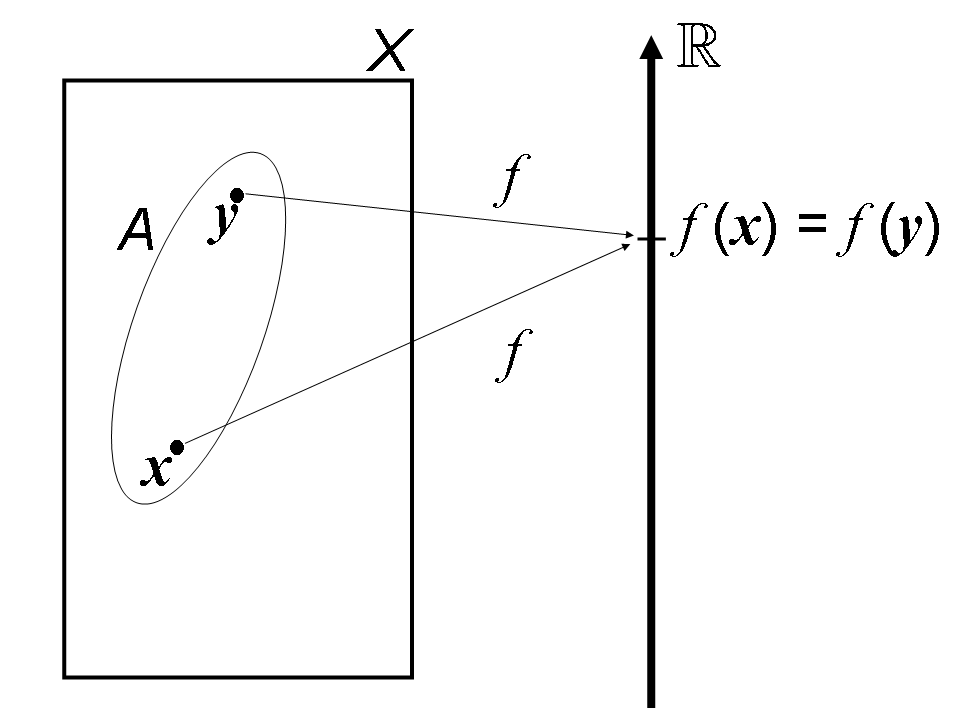}
\caption{\label{GA:Fig1} States $\vec{x}, \vec{y}$ in a state space $X$ of a classical system (left) and the
real numbers as the range of a classical observable $f: X \to \mathbb{R}$ (right). Epistemically equivalent
states $\vec{x}, \vec{y} \in X$ belong to the same equivalence class $A \subset X$.}
\end{figure}

Figure \ref{GA:Fig1} displays a situation in which an observable $f$ is not injective, such that different states $\vec{x} \ne \vec{y} \in A \subset X$ lead to the same measurement result
\begin{equation}\label{GA:Eq:epequiv}
f(\vec{x}) \;=\; f(\vec{y}) \quad .
\end{equation}
In this case, the states $\vec{x}$ and $\vec{y}$ are \emph{epistemically} indistinguishable by means of the observable $f$ (Shalizi and Moore 2003, beim Graben and Atmanspacher 2006).  Measuring $f$ cannot tell us whether the system is in state $\vec{x}$ or $\vec{y}$. The two states are \emph{epistemically equivalent} with respect to $f$ (beim Graben and Atmanspacher 2006).

In this way, the observable $f$ induces an equivalence relation ``$\sim_f$'' on the state space $X$: $\vec{x} \sim_f \vec{y}$ if $f(\vec{x}) = f(\vec{y})$. The resulting equivalence classes of ontic states partition the state space into mutually exclusive and jointly exhaustive subsets $A_1, A_2, \dots $ such that $A_i \cap A_j = \emptyset$ for all $i \ne j$ and $\bigcup_i A_i = X$. These subsets can be identified with the \emph{epistemic states} that are induced by the observable $f$.  More generally, we refer to subsets $S \subset X$ in state space as to epistemic states.\footnote{Epistemic states are actually defined as distributions over measurable sets from a $\sigma$-algebra in measure theory (beim Graben and Atmanspacher 2006). For a simplified exposition, which captures the very basic ideas, set-theoretical concepts are sufficient (cf.~beim Graben and Atmanspacher 2009).}
The collection $\mathcal{F} = \{ A_1, A_2, \dots \}$ of epistemic states is a then state space \emph{partition}.

We call $f$ an \emph{epistemic observable} if the partition $\mathcal{F}$ is not the \emph{identity partition} $\mathcal{I}$ where every cell $A_k$ is a singleton set containing exactly one element $A_k = \{ \vec{x}_k \}$ (Shalizi and Moore 2003). In this limiting case, $f$ is injective and becomes an  ontic observable. In the opposite limit, there is only one cell covering the entire state space $X$, and epistemic observables are constant over $X$: $f(\vec{x}) = \mbox{const}$ for all $\vec{x} \in X$. In this case, all states are epistemically equivalent with one another and belong to the (same) equivalence class $X$ of the \emph{trivial partition} $\mathcal{T}$.

\begin{figure}[!t]
\center
 \subfigure[]{\includegraphics[height=3.4cm]{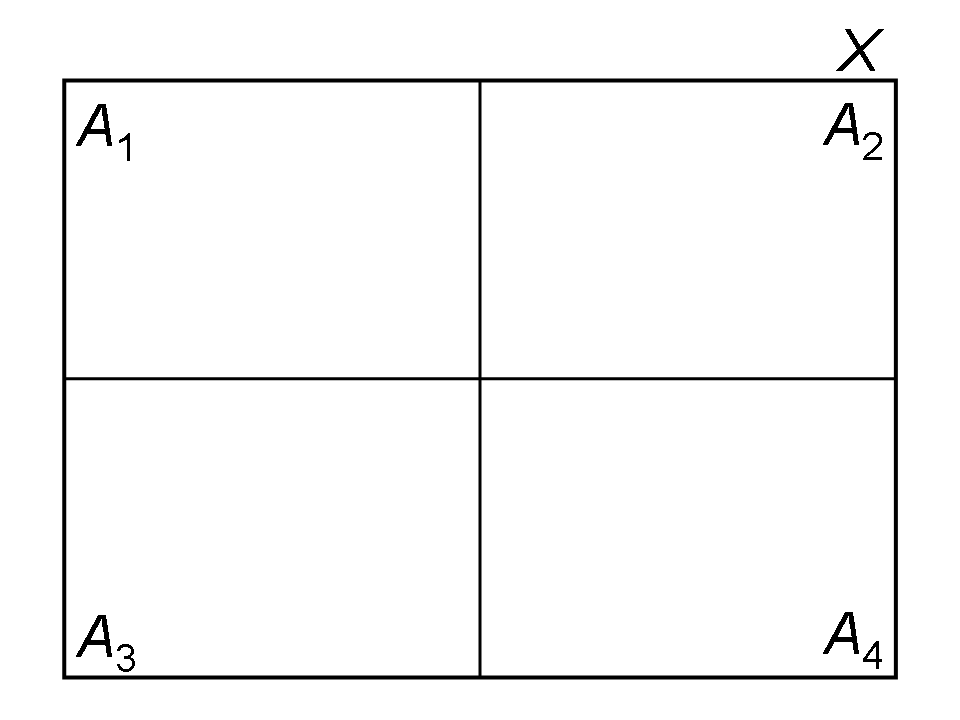}}
 \subfigure[]{\includegraphics[height=3.4cm]{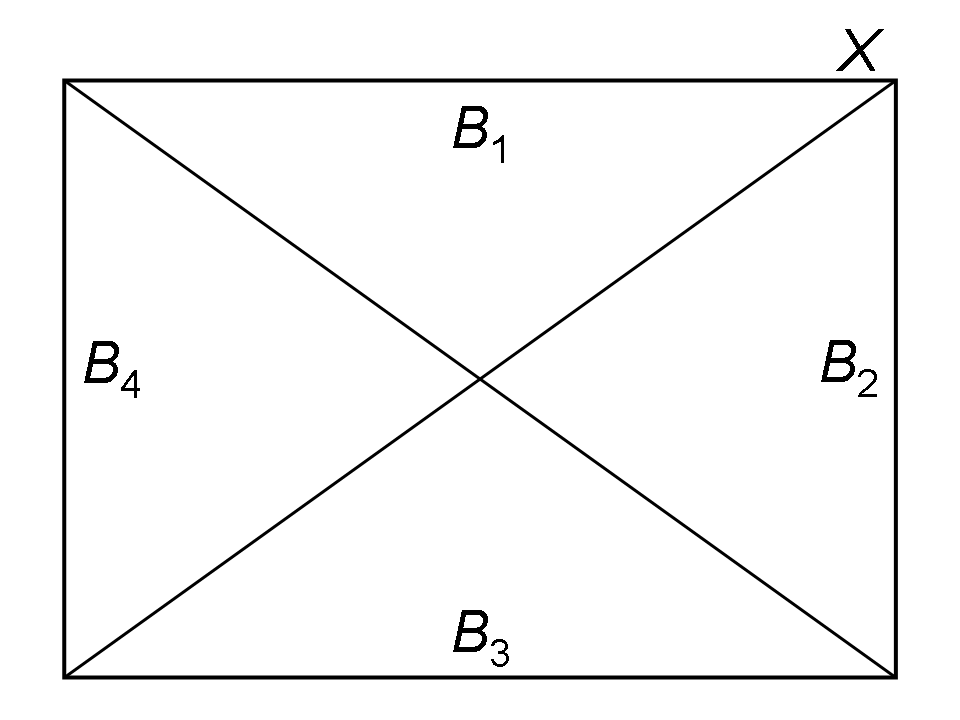}}
 \subfigure[]{\includegraphics[height=3.4cm]{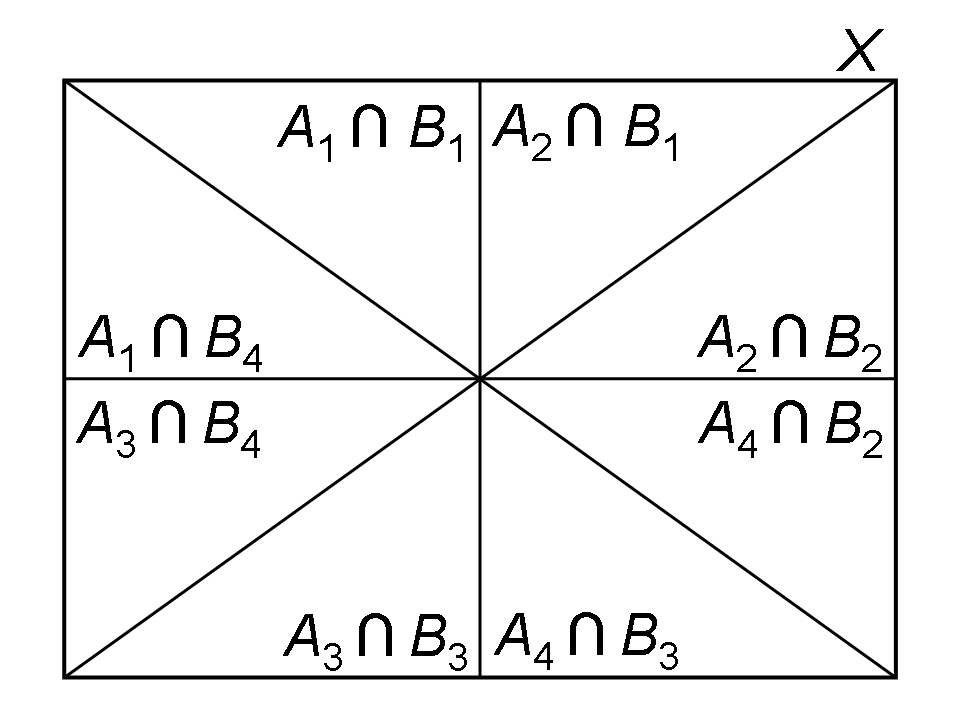}}
\caption{\label{GA:Fig2} Examples for finite partitions of the state space $X$: (a) ``rectangular'' partition
$\mathcal{F} = \{ A_1, A_2, A_3, A_4 \}$, (b) ``triangular'' partition $\mathcal{G} = \{ B_1, B_2, B_3, B_4 \}$, (c) product partition $\mathcal{F} \vee \mathcal{G}$.}
\end{figure}

Most interesting for our purposes are finite partitions $\mathcal{F} = \{ A_1, A_2, \dots A_n \}$ (where $n\in N$ is finite) which are neither trivial nor identity. Figures~\ref{GA:Fig2}(a,b) display two such finite partitions  $\mathcal{F}$ and $\mathcal{G}$ from which
a \emph{product
partition} $\mathcal{P} = \mathcal{F} \vee \mathcal{G}$ as in Figure 2(c) can be constructed. It contains all possible intersections of sets in $\mathcal{F}$ with sets in $\mathcal{G}$:
\begin{equation}\label{GA:Eq:partprod}
\mathcal{P} \;=\; \mathcal{F} \vee \mathcal{G} \;=\; \{ A_i \cap B_j | A_i  \in \mathcal{F}, B_j  \in \mathcal{G}\}\quad.
\end{equation}
The product partition $\mathcal{P}$ is a \emph{refinement} of both partitions $\mathcal{F}$ and $\mathcal{G}$. The refinement relation introduces a partial ordering relation ``$\prec$'' among partitions. If $\mathcal{G}$ is a refinement of $\mathcal{F}$, $\mathcal{G} \prec \mathcal{F}$, then there is a ``factor partition'' $\mathcal{H}$ such that $\mathcal{G} = \mathcal{F} \vee \mathcal{H}$. If neither $\mathcal{G}$ is a refinement of $\mathcal{F}$ nor {\it vice versa} (and $\mathcal{G} \ne \mathcal{F}$), the partitions $\mathcal{G}$ and $\mathcal{F}$ are called \emph{incomparable} (Shalizi and Moore 2003).

\subsection{Dynamics}
\label{dyna}
\begin{figure}[!t]
\center
\includegraphics[height=4cm]{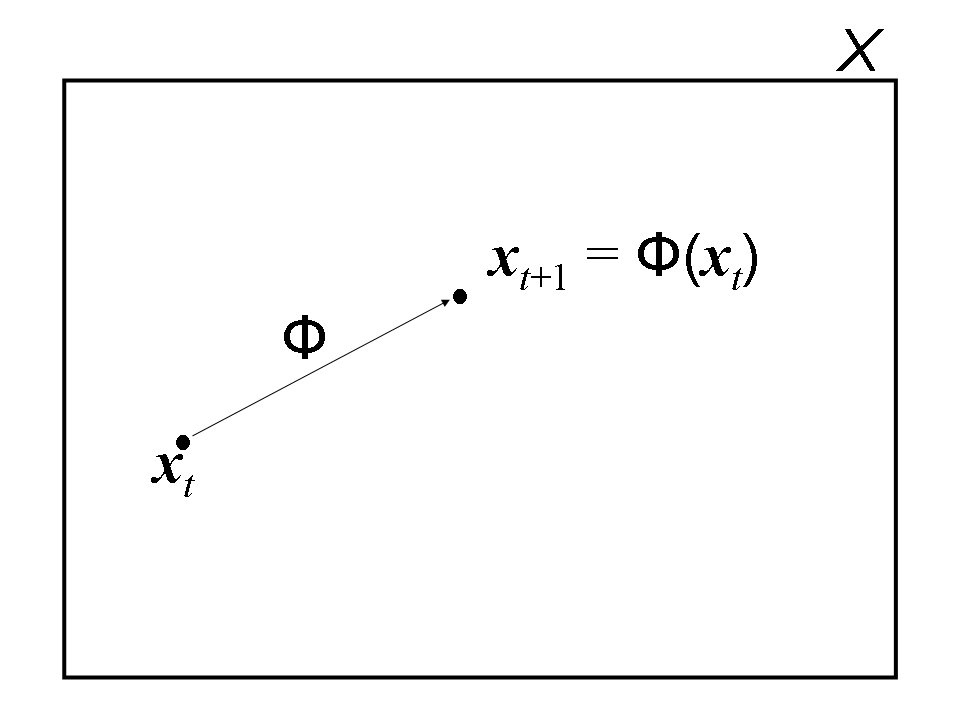}
\caption{\label{GA:Fig3} A discrete-time dynamics of a classical system is given by a map $\Phi: X \to X$ which assigns to a state $\vec{x}_t$ at time $t$ its successor
$\vec{x}_{t+1} = \Phi(\vec{x}_t)$ at time $t+1$.}
\end{figure}

A dynamical system evolves as a function of parameter time $t$. In other words, any present state  (e.g.~an initial condition) in state space, $\vec{x}_0 \in X$, gives rise to future states $\vec{x}_t \in X$. This evolution is described by a flow map $\Phi^t: X \to X$. In the simple case of a deterministic dynamics in discrete time, $\Phi$ maps any state $\vec{x}_t$ onto its successor $\vec{x}_{t+1}$, as illustrated in Fig.~\ref{GA:Fig3}.
Iterating the map $\Phi$ yields a \emph{trajectory}
\begin{equation}\label{GA:Eq:trajectory}
\vec{x}_{t+1} \;=\; \Phi^{t+1}(\vec{x}_0) \;=\; \Phi(\Phi^t(\vec{x}_0)) \;=\; \Phi(\vec{x}_t)
\end{equation}
for integer positive times $t \in \mathbb{N}$. Likewise, the inverse map $\Phi^{-1}$ can be iterated if the dynamics is invertible:
$\vec{x}_{-(t+1)} = \Phi^{-(t+1)}(\vec{x}_0) = \Phi^{-1}
(\Phi^{-t}(\vec{x}_0)) =\Phi^{-1}(\vec{x}_{-t})$, again for integer positive times $t \in \mathbb{N}$. In this way, the dynamics of an
invertible discrete-time system is described by the one-parameter group of integer numbers $t \in\mathbb{Z}$.
\subsection{Extended Measurements}
\label{cont}
In Sect.~\ref{stateobs}, we have described {\it instantaneous} measurements by the action of an observable $f: X \to\mathbb{R}$ on an ontic state
$\vec{x}$. Now we are able to describe \emph{extended
measurements}\footnote
{The notion of an extended measurement refers to a series of measurements extending over time $t$.}
by combining the action of an observable $f$ with the dynamics $\Phi$. Let the system be in state $\vec{x}_0\in X$ at time $t=0$. Measuring $f(\vec{x}_0)$ tells us to which class of epistemically equivalent states in the partition $\mathcal{F}$, associated with $f$, the state $\vec{x}_0$ belongs. Suppose that this is the cell
$A_{i_0} \in \mathcal{F}$. Suppose further that measuring $f$ in the subsequent state $\vec{x}_1 =\Phi(\vec{x}_0) \in X$ reveals that
$\vec{x}_1$ is contained in another cell $A_{i_1} \in \mathcal{F}$.

An alternative way to describe this situation is to say that the initial state $\vec{x}_0 =\Phi^{-1}(\vec{x}_1)$ belongs to the pre-image
$\Phi^{-1}(A_{i_1})$ of $A_{i_1}$. The information about $\vec{x}_0$ that is gained by measuring $f(\vec{x}_1)$ is, then, that the initial state $\vec{x}_0$ was contained in the intersection $A_{i_0} \cap \Phi^{-1}(A_{i_1})$. Continuing the observation of the system over one more instant in time yields that the initial state $\vec{x}_0$ belonged to the set $A_{i_0} \cap\Phi^{-1}(A_{i_1}) \cap \Phi^{-2}(A_{i_2})$ if the third measurement result was $\vec{x}_2 =\Phi^{2}(\vec{x}_0) \in A_{i_2}$.

\begin{figure}[!t]
\center
 \subfigure[]{\includegraphics[height=3.4cm]{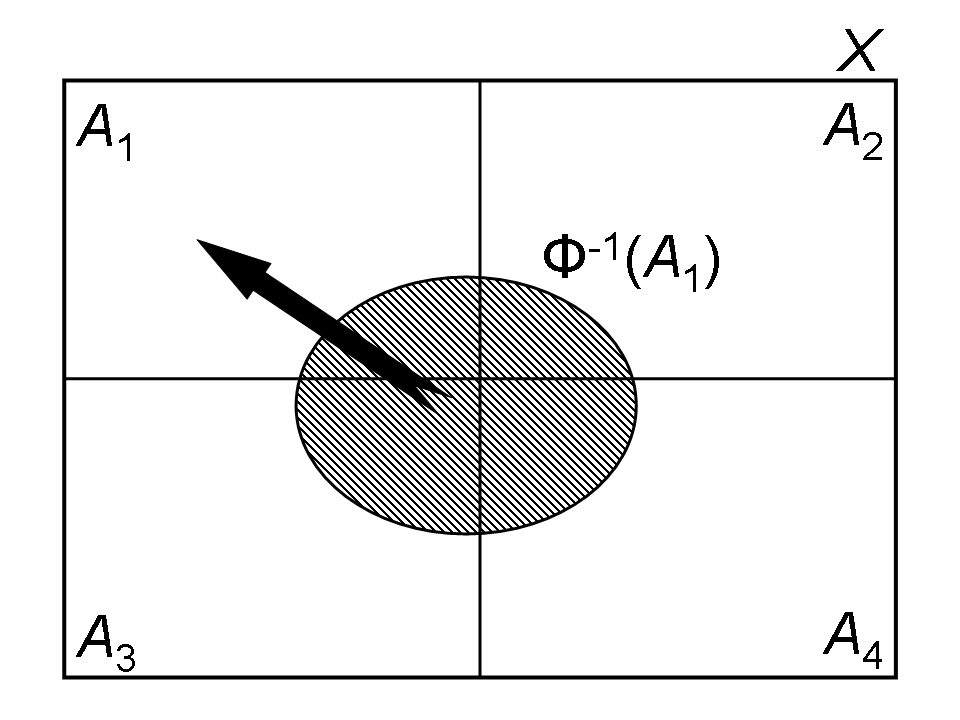}}
 \subfigure[]{\includegraphics[height=3.4cm]{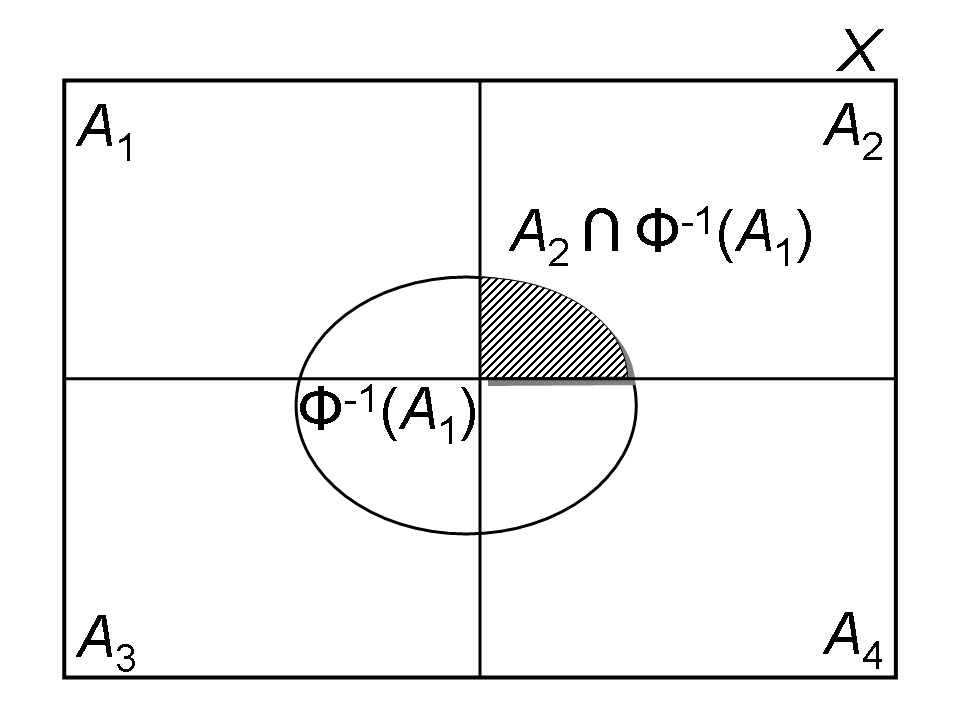}}
\caption{\label{GA:Fig4} Dynamic refinement of a partition. (a) For each cell $A_i$ of the partition $\mathcal{F}$ the pre-image
$\Phi^{-1}(A_i)$ under the dynamics is determined. The bold arrow indicates that the shaded region in state space is mapped onto cell $A_1$. (b) The shaded region in the product partition $\mathcal{F} \vee \Phi^{-1}(\mathcal{F})$ is the element $A_2 \cap \Phi^{-1}(A_1)$ of the dynamically refined partition.}
\end{figure}

A systematic investigation of extended measurements can now be based on the definition of the pre-image of a partition,
\begin{equation}\label{GA:Eq:preimage}
\Phi^{-1}(\mathcal{F}) \;=\; \{ \Phi^{-1}(A_i) | A_i \in \mathcal{F} \} \quad,
\end{equation}
which consists of all pre-images of the cells $A_i$ of the partition
$\mathcal{F}$. Then, an extended measurement over two successive time steps is defined by the product partition
$\mathcal{F} \vee\Phi^{-1}(\mathcal{F})$, containing all intersections of cells of the original partition $\mathcal{F}$ with cells of its pre-image $\Phi^{-1}(\mathcal{F})$. The result of the measurement of $f$ over two time steps is
$\vec{x}_0 \in A_{i_0} \cap \Phi^{-1}(A_{i_1}) \subset \mathcal{F} \vee \Phi^{-1}(\mathcal{F})$. This product partition is called the \emph{dynamic refinement} of $\mathcal{F}$, illustrated in Fig.~\ref{GA:Fig4}.

Most information about the state of a system can be gained by an ideal, ``ever-lasting'' extended measurement that began in the infinite past and will terminate in the infinite future. This leads to the \emph{finest dynamic refinement}
\begin{equation}\label{GA:Eq:finestrefine}
\mathbf{R}\,\mathcal{F} \;=\; \bigvee_{t = -\infty}^\infty \Phi^t(\mathcal{F}) \quad ,
\end{equation}
expressed by the action of the ``finest-refinement operator'' $\mathbf{R}$ upon a partition $\mathcal{F}$. It would be desirable that such an ever-lasting measurement yields complete information about the initial condition $\vec{x}_0$ in state space. This is achieved if the refinement (\ref{GA:Eq:finestrefine}) yields the identity partition,
\begin{equation}\label{GA:Eq:generator}
\mathbf{R}\, \mathcal{F} = \mathcal{I} \quad.
\end{equation}
A partition $\mathcal{F}$ obeying (\ref{GA:Eq:generator}) is called \emph{generating}. Generating partitions are structurally stable in the sense that they are robust under the dynamics. In other words, points on bondaries of cells are typically mapped onto points on boundaries of cells. In this way the epistemic states defined by the cells do not change over time --- which would be the case for non-generating partitions. For more details concerning the issue of stability in dynamical systems see Atmanspacher and beim Graben (2007).

Given the ideal finest refinement $\mathbf{R}\,\mathcal{F} = \mathcal{P}$ of a (generating or non-generating) partition $\mathcal{F}$ that is induced by an epistemic observable $f$, we are able to regain a description of extended measurements of arbitrary finite duration by joining subsets of $\mathcal{P}$ which are visited by the system's trajectory. Supplementing the ``join'' operation by the other Boolean set operations over $\mathcal{P}$ leads to a \emph{partition algebra} $A(\mathcal{P})$ of $\mathcal{P}$. Then, every set in $A(\mathcal{P})$ is an epistemic state measurable by $f$.

Note that the concept of a generating partition in the ergodic theory of deterministic systems is related to the concept of a {\it Markov chain} in the theory of stochastic systems. Every deterministic system of first order gives rise to a Markov chain which is generally neither ergodic nor irreducible. Such Markov chains can be obtained by so-called \emph{Markov partitions} that exist for expanding or hyperbolic dynamical systems (Sinai 1968, Bowen 1970, Ruelle 1989). For non-hyperbolic systems no corresponding existence theorem is available, and the construction can be even more tedious than for hyperbolic systems (Viana \emph{et al.} 2003). For instance, both Markov and generating partitions for nonlinear systems are generally non-homogeneous. In contrast to Figure~\ref{GA:Fig2}, their cells are typically of different size and form.

Note further that every Markov partition is generating, but the converse is not necessarily true (Crutchfield 1983, Crutchfield and Packard 1983). For the construction of ``optimal'' partitions from empirical data it is often more convenient to approximate them by Markov partitions (Froyland 2001; see also Deuflhard and Weber 2005, Gaveau and Schulman 2005). See Allefeld {\it et al.}~(2009) for a concrete example of how a Markov partition can be constructed from empirical data.

\section{Compatibility and Complementarity in Classical Dynamical Systems}
\label{cccdyn}
If a partition $\mathcal{F}$ is not generating, its finest refinement is not the identity partition. In this case, the refinement operator yields a partition $\mathcal{P} = \mathbf{R} \mathcal{F}$ with some residual coarse grain. Moreover, the cells of a non-generating partition are not stable under the dynamics $\Phi$, so that they become dynamically ill-defined --- a disaster for any attempt to formulate a robust coarse-grained (epistemic) description (Bollt {\it et al.} 2001, Atmanspacher and beim Graben 2007).

Let $P \in \mathcal{P}$ be an epistemic state of the finest refinement of $\mathcal{F}$. Because $\mathcal{F}$ is induced by an observable $f$ whose epistemic equivalence classes are the cells of $\mathcal{F}$, all cells of $\mathcal{P}$ can be accessed by extended measurements of $f$. However, as $\mathcal{P}$ is not the identity partition $\mathcal{I}$, the singleton sets $\{ \vec{x} \}$ representing ontic states in $X$ are not accessible by measuring $f$. An arbitrary epistemic state $S \subset X$ is called \emph{epistemically accessible with respect to $f$} (beim Graben and Atmanspacher 2006) if $S$ belongs to the partition algebra $A(\mathcal{P})$ produced by the finest refinement of $\mathcal{F}$.

Measuring the observable $f$ in all ontic states $\vec{x} \in P$ belonging to an epistemic state $P \in\mathcal{P}$ always yields the same result $a = f(\vec{x})$ since $f$ is by construction constant over $P$. Therefore, the variance of $f(\vec{x})$ across $P$ (with respect to some probability measure) vanishes such that $f$ is dispersion-free in the epistemic state $P$. In other words, $P$ is an eigenstate of $f$. One can now easily construct another observable $g$ that is not dispersion-free in $P$ such that $P$ is not a common eigenstate of $f$ and $g$. As a consequence, the observables $f$ and $g$ are incompatible as they do not share all (epistemically accessible) eigenstates. Beim Graben and Atmanspacher (2006) refer to this construction as an \emph{epistemic quantization} of a classical dynamical system.

In an ontic description of a classical system, ontic states are common eigenstates of all observables. Therefore, classical observables associated with ontic states are always compatible. By contrast, if the ontic states are not epistemically accessible by extended measurements, the smallest epistemically accessible states are cells in the finest refinement of a partition $\mathcal{F}$ induced by an epistemic observable $f$. These epistemic states are not eigenstates of every observable, such that observables associated with them are incompatible. As in quantum theory, two observables $f$ and $g$ are complementary if they do not have any (epistemically accessible) eigenstate in common, i.e.~if they are maximally incompatible (Raggio and Rieckers 1983). Beim Graben \emph{et al.} (2013) demonstrated the incompatibility of position and momentum of a classical harmonic oscillator subjected to time-discretization and spatial coarse-graining.

Nevertheless, even in an epistemic description, classical observables $f$ and $g$ can be compatible with one another. This is the case if all ontic states $\vec{x} \in X$ are epistemically accessible with respect to both $f$ and $g$. The necessary and sufficient condition for this is that the partitions $\mathcal{F}$, $\mathcal{G}$ be generating (Eq.~6). This leads to a generalization of the concepts of compatibility and complementarity: Two partitions $\mathcal{F}, \mathcal{G}$ are called compatible if and only if they are both generating:
$\mathbf{R}\,\mathcal{F} = \mathbf{R}\,\mathcal{G} = \mathcal{I}$. They are incompatible if $\mathbf{R}\,\mathcal{F} \ne\mathbf{R}\,\mathcal{G}$, which is always the case if at least one partition is not generating. They are complementary if their finest refinements are disjoint:
$\mathbf{R}\,\mathcal{F} \cap\mathbf{R}\,\mathcal{G} = \emptyset$.

\section{Non-Boolean Logic of Propositions}

Already in their seminal paper on quantum logics, Birkhoff and von Neumann (1936) asked for the propositional calculus emerging from the epistemic restrictions upon an arbitrarily selected observation space.  A proposition such as ``the observable $f$ assumes the value $a$ in state $\vec{x} \in X$'', or briefly ``$a = f(\vec{x})$'', induces a binary partition of the state space $X$ of a classical dynamical system into two subsets of
\begin{equation}\label{binpart}
    \mathcal{F} = \{ S, \: X \setminus S \} \,,
\end{equation}
where $S = \{ \vec{x} \in X | a = f(\vec{x}) \}$. Because propositions can be combined by the logical
connectives ``and'', ``or'', and ``not'', the structure of a \emph{classical propositional logic} is that of a Boolean
algebra of subsets of the state space (Birkhoff and von Neumann 1936, Primas 1977, Westmoreland and Schumacher 1993).

Given a classical dynamical system with state space $X$, dynamics $\Phi$, and a representatively chosen epistemic observable $f$, this observable induces a partition $\mathcal{F}$ whose finest refinement is $\mathbf{R} \mathcal{F}$. The partition algebra $A(\mathbf{R} \mathcal{F})$, comprising all subsets of $X$ that can be formed by the Boolean set operations of intersection, union, and negation, applied to the states that are epistemically accessible by means of extended measurements of $f$, is a Boolean set algebra describing a classical propositional logic.

However, things become much more involved when we add another epistemic observable $g$, that is not compatible with $f$. In that case, the finest refinements $\mathbf{R}\,\mathcal{F} \ne\mathbf{R}\,\mathcal{G}$ yields  overlapping partition algebras $A(\mathbf{R} \mathcal{F}) \cap A(\mathbf{R} \mathcal{G}) \ne \emptyset$. If the overlap is not trivial (i.e. neither $\emptyset$ nor $X$), these partition algebras form a \emph{partition logic}, or equivalently a \emph{partition test space} (Dvure\v{c}enskij \emph{et al.} 1995).

Test spaces have been introduced as \emph{generalized sample spaces} in a so-called \emph{operational statistics} (Foulis and Randal 1972, Randal and Foulis 1973) in order to clarify problems of incompatible observables in general (classical) measurement situations. Their argumentation is nicely illustrated by Foulis' (1999) ``firefly box'' thought experiment:

Imagine a firefly caught in a box with only two translucent windows, one at the front and one at one side of the box. Assume that both windows allow us to assess the firefly's position only in a coarse-grained manner: either left (``L'') or right (``R'') with respect to the front view, and front (``F'') or back (``B'') with respect to the side view. Because the firefly can be observed either from the front or from the side, these ``measurements'' are mutually exclusive and can give rise to incompatible descriptions.

\begin{figure}[!t]
\center
\includegraphics[height=6cm]{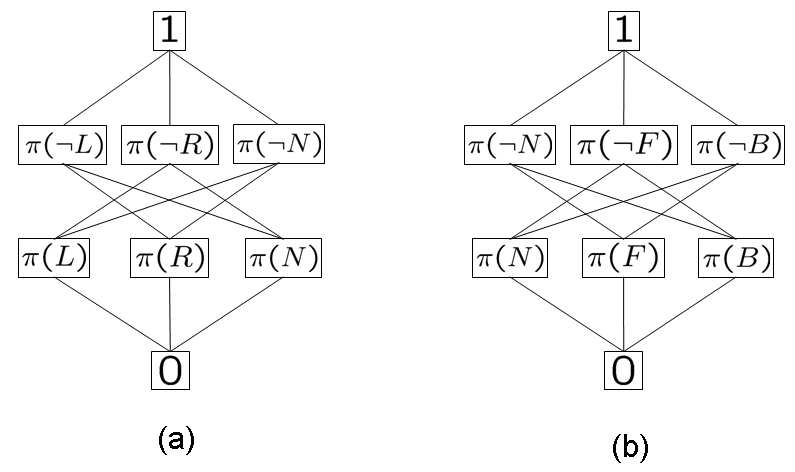}
\caption{\label{GA:Fig5} Boolean propositional lattices for two observations of the firefly in the box: (a) front view, (b) side view.}
\end{figure}

That this is in fact the case results from a third possibility: the firefly does not glow (``N''), which creates epistemically equivalent events that are subsumed in the following six propositions: (1) firefly is somewhere, (0) firefly is nowhere, $\pi(L)$ firefly is in the left, $\pi(R)$ firefly is in the right, $\pi(N)$ firefly is not glowing, $\pi(F)$ firefly is in the front, $\pi(B)$ firefly is in the back, and the respective negations thereof (indicated by ``$\neg$''). Figure \ref{GA:Fig5} displays the lattice diagrams of the resulting Boolean logics.
\begin{figure}[!b]
\center
\includegraphics[height=7cm]{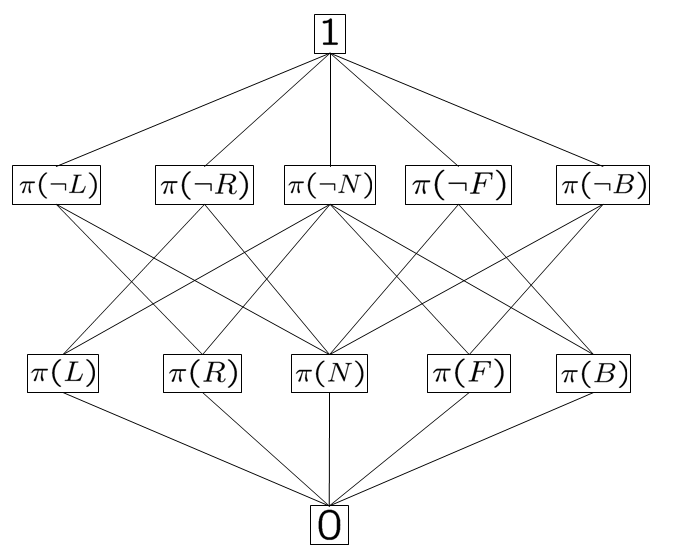}
\caption{\label{GA:Fig6} Non-Boolean (orthomodular) propositional lattice for joint observations of the firefly in the box.}
\end{figure}

Crucially, the Boolean lattices depicted in Figures \ref{GA:Fig5}(a, b) can be pasted together along their common overlaps $0, 1, N, \neg N$, giving rise to the lattice shown in Fig. \ref{GA:Fig6}. By this procedure one obtains a so-called \emph{orthomodular lattice} (Primas 1977). According to Piron's representation theorem (and under some technical requirements, see Baltag and Smets 1995, and Blutner and beim Graben to appear), orthomodular lattices can be represented by the orthomodular projector lattice of a Hilbert space. Hence, the firefly example indicates how nontrivially overlapping partition algebras induced by incompatible observables can be pasted together to orthomodular lattices forming the basis of non-Boolean logics (Dvure\v{c}enskij \emph{et al.} 1995, Greechie 1968, Blutner and beim Graben to appear). The case of a non-Boolean lattice with Boolean sublattices is also called a partial Boolean lattice (Primas 2007 and references therein).

\section{Implications}

\subsection{Entanglement Beyond Quantum Physics}

The notion of complementarity and the non-Boolean structure of quantum logic are tightly connected to a third basic feature of quantum theory: {\it entanglement}. In quantum physics proper, entanglement relies on the dispersion of ontic (pure) quantum states. While classical ontic (pure) states are dispersion-free and, thus, cannot be entangled, it is still possible to define a kind of epistemic entanglement for classical epistemic (mixed) states based on state space partitions.

This was proven, and illustrated by examples, in a recent paper by beim Graben {\it et al.} (2013). The strategy of the proof rests on properly defined ``epistemically pure'' dispersive classical states, in contrast to ``ontically pure'' dispersive quantum states. Epistemically pure classical states arise out of the finest refinement of a partition with respect to all epistemically realizable observables. We demonstrated that, under particular conditions, classical states that are epistemically pure with respect to a global observable may be mixed with respect to a suitably defined local observable.

The structure of this argument resembles the definition of quantum entanglement, where the pure state of a system as a whole is not separable into pure states of decomposed subsystems. The decomposition leads to mixed states with nonlocal correlations between the subsystems. It can be demonstrated how a related inseparability occurs for classical dynamical systems with non-generating partitions (see beim Graben {\it et al.} 2013).

Nonlocal correlations in this case are due to partition cells whose boundaries change dynamically and lead to non-empty intersections among cells. These correlations contribute to the dynamical entropy of the system and imply an underestimation of its Kolmogorov-Sinai entropy.\footnote{The Kolmogorov-Sinai entropy of a dynamical system is defined as the supremum of its dynamical entropy over all partitions. This supremum is attained by a generating partition.} Allahverdijan {\it et al.} (2005) studied such a case for Brownian motion: They used a homogeneous partition with respect to particle velocities to determine correlations in the system considered. Such a partition is not generating, which raises the overall amount of correlations and can lead to the impression of entanglement. This is consistent with their observation that increasing refinement of the partition entails decreasing correlations.

Our results substantiate a recent conjecture of epistemic entanglement by Atmanspacher {\it et al.}
(2011). It should be of interest to all kinds of applications which can be formalized in terms of state space represenations. This applies in particular to situations in cognitive science for which state space representations have generated increasing attention. Recent work by Allefeld {\it et al.} (2009) shows how generating state space partitions can be constructed so as to avoid quantum features in mental systems.

\subsection{Quantum Mind Without Quantum Brain}

An important criterion for entanglement correlations is the violation of Bell-type inequalities.
There is considerable recent interest in various kinds of bounds inherent in such inequalities as applied to mental systems (Atmanspacher and Filk 2014, Dzhafarov and Kujala 2013). One thrilling application is due to Bruza {\it et al.} (2009), who challenged a long-standing dogma in linguistics by proposing non-separable concept combinations in the human mental lexicon. Another intriguing example is due to Atmanspacher and Filk (2010) who suggested temporally nonlocal phenomena in mental states.

Temporally nonlocal mental states can be interpreted as states that are not sharply (pointwise) localized along the time axis, and their characterization by sharp (classical) observables is inappropriate. Rather, such states appear to be ``stretched'' over an extended time interval whose length may depend on the specific system considered. Within this interval, relations such as ``earlier'' or ``later'' are illegitimate designators and, accordingly, causal connections are ill-defined.

There is accumulating evidence that concepts like complementarity, entanglement, dispersive states, and non-Boolean logic play significant roles in mental processes (see Wang {\it et al.} (2013) for a compact recent overview). Within the traditional framework of thinking, this would imply that the brain activity correlated with those mental processes is in fact governed by quantum physics. Quite a number of quantum approaches to consciousness have been proposed along these lines (cf.~Atmanspacher 2011), with little empirical support so far.

Our results underline that quantum brain dynamics is not the only possible explanation of quantum features in mental systems. Assuming that mental states arise from partitions of neural states in such a way that statistical neural states are co-extensive with individual mental states, the nature of mental processes depends strongly on the kind of partition chosen. If the partition is not generating, it is possible or even likely that mental states and observables show features that resemble quantum behavior although the correlated brain activity may be entirely classical --- quantum mind without quantum brain.

\section*{Acknowledgments}

Large parts of the material in Sections 2 and 3 are reproduced from Sections 3 and 4 in beim Graben and Atmanspacher (2009), which the present paper refines and develops.

\end{document}